\documentclass[letterpaper, paper,11pt]{AAS}		

\usepackage{bm}
\usepackage{amsmath}
\usepackage{subfigure}
\usepackage[colorlinks=true, pdfstartview=FitV, linkcolor=black, citecolor= black, urlcolor= black]{hyperref}
\usepackage{footnpag}			      	
\usepackage{epstopdf}
\usepackage{epsfig}
\usepackage{algorithm}
\usepackage{algpseudocode}
\usepackage{pifont}
\usepackage{gensymb}
\usepackage{overcite}
\usepackage{hyperref}

\PaperNumber{15-524}

\begin{document}

\title{Orbit Determination and Differential-Drag Control of Planet Labs Cubesat Constellations}

\author{Cyrus Foster\thanks{Orbit Mechanic, Planet Labs, cyrus@planet.com},  
Henry Hallam\thanks{Chief Astronaut, Planet Labs, henry@planet.com},
\ and James Mason\thanks{Director of Missions, Planet Labs, james@planet.com}
}

\maketitle{}

\begin{abstract}
We present methodology and mission results from orbit determination of Planet Labs nanosatellites and 
differential-drag control of their relative motion.  Orbit determination (OD) is required on Planet Labs 
satellites to accurately predict the positioning of satellites during downlink passes and we present a 
scalable OD solution for large fleets of small satellites utilizing two-way ranging.  
In the second part of this paper, we present mission results from relative motion differential-drag 
control of a constellation of satellites deployed in the same orbit.
\end{abstract}

\section{Introduction}

Planet Labs Inc.  is a purpose-driven space and data analytics company based in San Francisco, California.  
The company designs, builds and operates large fleets of Earth observation nanosatellites 
to image the entire planet at an unprecedented frequency.  
Planet Labs aims to provide universal access to information about the changing planet 
to enable both commercial and humanitarian applications.  

Planet Labs' concept of operations differs fundamentally from other optical remote sensing satellites
in that the spacecraft always point their cameras towards nadir and are “always on”.  
Satellites always image while over land with reasonable solar illumination, and tasking is only done at the downlink stage.  
A single orbital plane of nadir-pointing satellites can image the whole Earth every single day 
if there are enough of them that their imaging swaths overlap.  
For the design altitude and telescope field of view of our instrument, 
a minimum of 105 evenly-spaced satellites is required to ensure daily coverage as the Earth rotates under the orbital plane.  
Evenly spaced satellites additionally allow for the optimal use of the limited number of ground stations available 
for downloading all of the imagery data.  
For these reasons, station keeping is strongly preferred over random distributions of satellites along the orbit.  

Operating fleets of satellites this size presents unique challenges, 
not least of which are orbit determination and station keeping control.
The necessity to track each satellite accurately, and the ability to deploy, disperse, and then station-keep 
is critical to the optimal operation of the fleet.  
These capabilities were identified early on in the design of Planet Labs’ concept of operations, 
and the ability to station-keep without the complexity and expense of onboard propulsion 
drove us to develop in-house differential drag control technology.  
This paper presents a status update on the significant investment made over the last four years towards solving this problem.

On orbit determination, we present an introduction to our UHF radio ranging capability, 
and the resulting orbit determination capability that allows us to track our satellites 
independently of, and more accurately than third party service such as JSpOCs.  

On differential drag, we present the basic physical concepts, 
the enabling attitude configurations, our constellation deployment strategy, 
and the methodology and mission results from differential-drag control of their relative motion.

\section{Orbit Determination}

Orbit determination (OD) is required on Planet Labs satellites for
multiple operational purposes including on-orbit attitude control,
image capture and downlink scheduling, georectification of payload
imagery, conjunction assessment, and most critically, pointing of
ground station antennas during downlink passes.  Imagery is downloaded
from a satellite via a high-bandwidth X-band link requiring ground
station antenna pointing errors of less than $0.2$\degree.  An
ephemeris for the satellite must be known in advance of a downlink
pass to compute an appropriate Two-line element set (TLE) or
azimuth-elevation file for a specific pass.  TLEs published by the
Joint Space Operations Center (JSpOC) are not accurate or reliable
enough for nominal operations, especially for satellites in low orbits
(\textless $450$ km) where atmospheric and spacecraft attitude
uncertainties yield large perturbations with orbit propagation.  JSpOC
spacecraft identification cross-tagging issues have also been
encountered as a consequence of a large co-deployed fleet of small
spacecraft.

We present a scalable OD solution for large fleets of small
satellites.  Two-way time-of-flight ranging measurements are taken at
UHF ground stations and are subsequently fit to an orbit using a
high-precision propagator with a numerical force model.  Orbits are
then propagated into the future to produce a TLE or pass-specific
azimuth-elevation files.  We present the methodology for fitting an
orbit to this data and mission results.

The satellites, or “Doves”, fall into two orbit-regimes:
sun-synchronous inclination at $600$ km altitude and sub-ISS orbits
(\textless$400$ km at $51.6$\degree inclination).  At publication
date, Planet Labs currently has 12 satellites in the sun-synchronous
orbit, and 25 satellites in the sub-ISS orbit.

\subsection{Quality of JSpOC TLEs}

Especially when deployed in large numbers into low-altitude orbits
(e.g.  via release from the International Space Station), cubesats can
present a challenging problem\cite{kat} to JSpOC or any other entity
attempting to track them via optical and radar means.  The combination
of small radar cross-section and rapid evolution in the orbit due to
varying atmospheric density results in sparse, noisy measurements that
must be correlated and tagged by JSpOC's semi-automated systems.
Because each individual measurement provides no unambiguous
identification of which object it represents, some cross-tagging
inevitably occurs, leading to misidentification in the Satellite
Catalog and/or corrupted solutions (when measurements from more than
one object are mistakenly included in the same orbit fit).  Accuracy
of the TLEs published on Space Track\footnote{\url{https://www.space-track.org}} 
(as compared to truth from RF and
laser ranging, GPS and phased-array imaging) has been highly variable
and difficult to predict.  Over the course of a few months, position
accuracy of the TLEs issued for a particular object may vary from
better than 1 km up to several hundred km of error.

\subsection{Ranging data acquisition}

The aforementioned TLE quality issues, discovered early in Planet
Labs' Flock 1 mission, accelerated development of an independent
means of orbit determination.  Replacement firmware for the
satellites' UHF Low Speed Transceiver (LST) was developed and uploaded
to the spacecraft already on-orbit, to allow two-way time-of-flight
ranging measurements to be made.  Unlike the high-bandwidth X-band
link, the LST can tolerate ground station pointing errors in excess of
$10\degree$, allowing bootstrapping from an initially coarse orbit
estimate.

The LST ranging packets are tagged with the serial number of the
satellite to be interrogated, ensuring positive identification and
eliminating the possibility of cross-tagging.  Ranging stations use
commodity Yagi-Uda antennas and positioners, and can be deployed
worldwide at very low cost.  Since many satellites may be
simultaneously in view from any site, a program was developed to
automatically decide which satellite to interrogate based on various
heuristics, including the satellites' range and elevation, angular
distances from current antenna position, distribution of recently
gathered ranging data and operator-assigned priority factors.

\begin{figure}[!ht]
\centering
\includegraphics[width=0.7\textwidth]{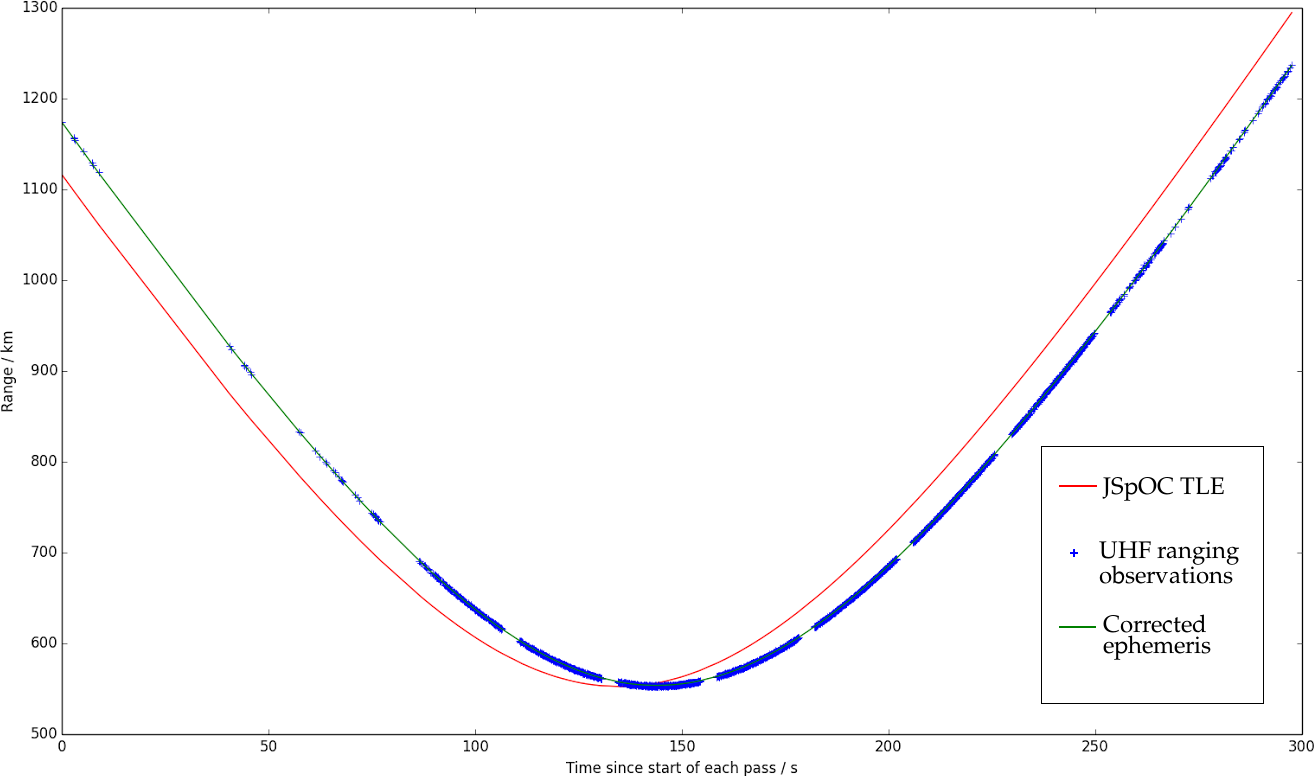}
\caption[caption]{Example of ranging data acquired during a pass vs.  JSpOC TLE}
\label{fig:ODpostprocess}
\end{figure}

Approximately 350,000 ranging measurements are gathered daily across
three dedicated ranging stations.  The $1\sigma$ accuracy of each
measurement is 650 m.  
Planet Labs recommends that all cubesat missions include some means of
independent orbit determination, and to facilitate this has released
an open-source version of the Low Speed Transceiver design.

\subsection{Orbit fitting to ranging data}

We fit an orbit to acquired ranging data using a high-precision propagator and least-squares solver.  
The propagator numerically integrates cartesian state vectors using the Adams-Moulton variable-order and variable-step method.
We use a 10x10 EGM96 spherical harmonic gravity field and the NRL-MSISE00 atmosphere model with latest space weather and 
Earth Orientation Parameters (EOP) data from Celestrak\footnote{\url{http://celestrak.com}}.

\begin{algorithm}
\caption{Orbit Determination Algorithm}
\label{ODalgorithm}
\begin{algorithmic}[1]
	\Procedure{Fit orbit given ranging observations}{}
	\State get state vector inital guess $X$ from previous run
	\State compute diagonal weight $W$ from sensor 1-sigmas: $W_{i,i}=\frac{1}{\sigma_{\text{ranging}}^2}$
	\While{state vector position update is large}
	\State propagate state vector to epochs of ranging observations
	\State compute partials matrix $A = \frac{\delta Range}{\delta X}$, 
	and residuals $\Delta y = \text{Range}_{\text{observed}} - \text{Range}_{\text{propagated}}$
	\State compute covariance matrix $P = (A^TWA)^{-1}$
	\State adjust fit span and fit variables as appropriate based on $P$ (on 1\textsuperscript{st} iteration only)
	\State update state vector with $\delta X = P A^T W \Delta y$
	\EndWhile
	\EndProcedure
\end{algorithmic}
\end{algorithm}

\begin{table}[htbp]
\fontsize{10}{10}\selectfont
\caption{State vector fit elements depend on quality of observations}
\label{FitElements}
\centering
\begin{tabular}{ccc}
Nominal case &
Poor observability &
Very poor observability \\
$X = \begin{bmatrix} r_x \\ r_y \\ r_z \\ v_x \\ v_y \\ v_z \\ \frac{1}{BC} \end{bmatrix}$
&
$X = \begin{bmatrix} r_R \\ r_S \\ v_R \\ v_S  \end{bmatrix}$
& 
$X = \begin{bmatrix} r_{alongtrack} \end{bmatrix}$
\end{tabular}
\end{table}

To fit a state vector to ranging observations, we make iterative corrections 
to an initial guess using a batch Gauss-Newton least-squares solver as described in Algorithm \ref{ODalgorithm} \cite{vallado}.
The initial guess is obtained from a state vector produced the last time it was solved for 
and propagated forward in time to the present epoch, 
but an approximate guess from a JSpOC TLE can also be used.  
While it should be possible to obtain an approximate state vector from only the ranging data 
via Initial Orbit Determination (IOD), to date there has never been a need to do so.

We decide which elements of the state vector to fit based on various criteria for the covariance matrix $P$.  
For the large majority of cases, we can fit the full state vector $X$, 
but in some cases we have to resort to only fitting a subset of state vector elements because of 
poor observability, such as when the satellite has recently been deployed 
and may have only had a single ranging pass (see table \ref{FitElements}).

For example, if the magnitude of the position covariance is too high, 
we fit position and velocity vectors components in the orbit plane (R and S unit vectors in the RSW satellite frame\cite{vallado}).
If the position covariance is still too high, we then resort to fitting the satellite's mean anomaly via an along-track proxy.
Similarly, the ballistic coefficient $BC$ is fit if its covariance is small enough,
its inclusion judged independently of the position and velocity components.
We are also careful not to use a fit span that is too long when the satellite is in a high-drag environment
to ensure partial derivatives sensitivities do not exhibit highly non-linear properties.
We fit the inverse $BC$ because it is numerically more convenient and exhibits more linear behaviour.

Similarly, we choose an appropriate span of ranging data based on position and $BC$ covariances from $P$.
Typically, fit spans of 3-4 days are selected to be appropriate for the 600 km sun-synchronous orbit, 
while a shorter span of 0.5-1 day is ideal for the high-drag sub-ISS orbits.

\subsection{GPS data}

Some of the satellites carry an experimental software-defined GPS
receiver, which due to power constraints and radio-frequency
interference cannot be operated continuously. However, the GPS can provide sparse,
highly precise position/velocity/time measurements against which
ranging data and determined orbits can be validated.

To produce a high-quality "truth" ephemeris, we take segments of time
that feature densely and consistently-acquired GPS observations.  We
then fit a state vector (including the ballistic coefficient) to all
available GPS observations within an appropriate fit span, and
sequentially repeat this process to cover the span of available GPS
data.  Finally, we evaluate position at desired times by interpolating
between states propagated from bounding fit epochs.

While the GPS data is not always available on the current generation
of satellites, making it unsuitable for routine orbit
determination, it is very useful as an independent source of
observations.  For example, the GPS data has allowed us to find and
account for biases in the ranging data.  We expect GPS to be
fully-functional on future satellite builds, providing a more
accurate, if higher latency, source of observations for automated
operational orbit determination.

\subsection{Orbit prediction results}

In figure \ref{fig:ODpostprocess}, we show position offsets for 
a ranging-derived ephemeris and JSpOC TLEs, as compared to a high-quality GPS-derived ephemeris.  
The ranging ephemeris is produced using all available ranging data, 
and we interpolate between bounding TLEs published during the period of interest to produce a JSpOC-derived ephemeris.
The comparison is made for satellite Flock 1c-5 (NORAD CATID 40038) at reference epoch 2015-02-24 14:00:00 UTC, 
during a 12-day period where GPS measurements were acquired regularly.
At the time, the satellite was in a $590$ x $620$ km altitude sun-synchronous orbit.

We observe a position offset for ranging-derived and JSpOC TLEs of 0.45 and 1.45 km RMS respectively 
(maximums of 0.86 and 3.46 km) when compared to the high-quality GPS ephemeris in the 24 hours after the reference epoch.
These results are representative of what we observe for Planet Labs satellites in sun-synchronous orbits.
Satellites in the lower sub-ISS orbit will also feature similar behaviour at times but can also exhibit
higher position offsets in the 10-20 km range during periods of inconsistent attitude maneuvers and space weather variability 
that make orbit fitting and propagation with a fixed ballistic coefficient less appropriate.  
These effects are especially pronounced as the satellite encounters an 
increasingly dense atmosphere as its altitude decays. 

\begin{figure}[!ht]
	\centering\includegraphics[width=4.5in]{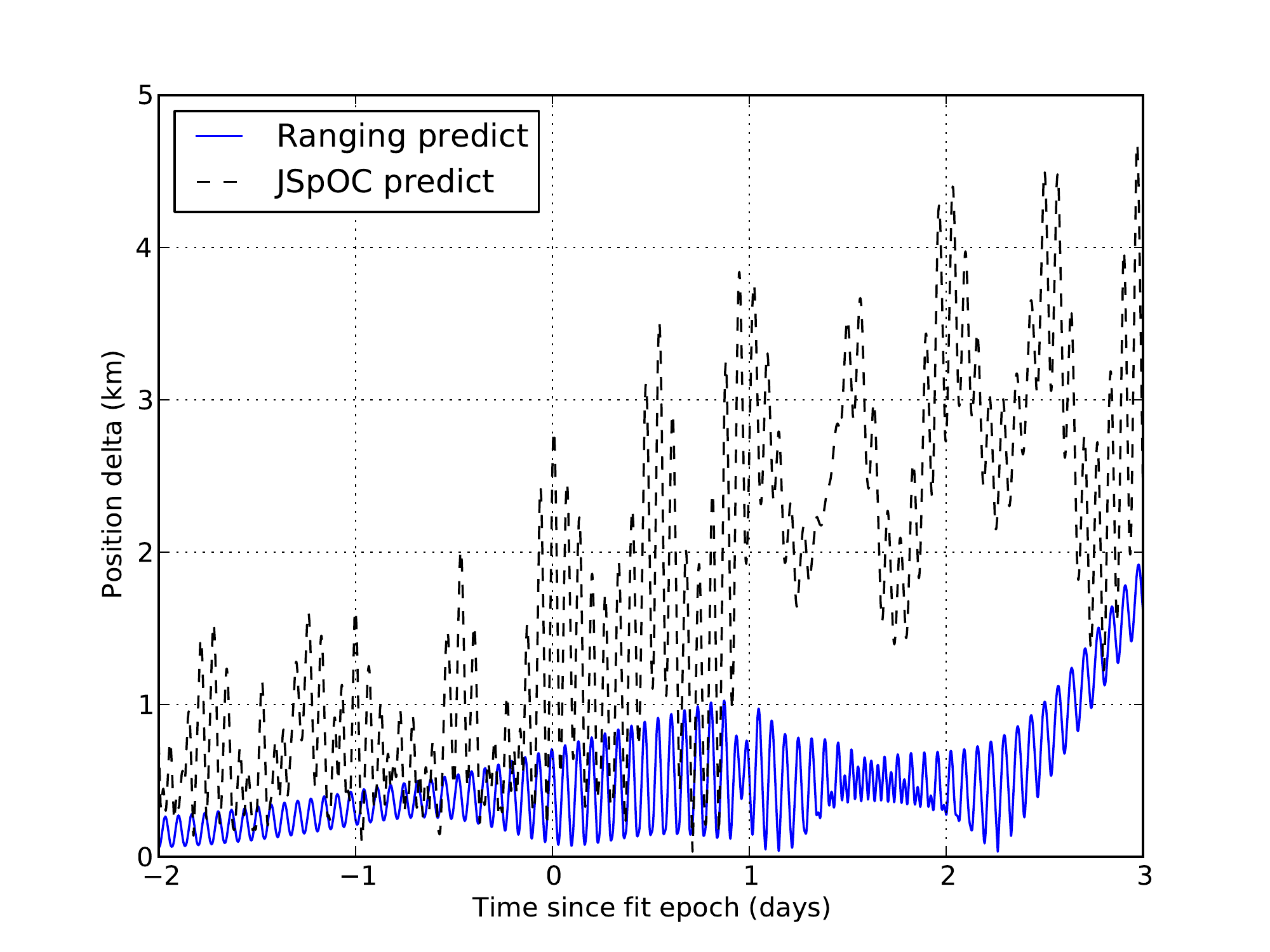}
	\caption[caption]{GPS Truth vs.  Predictions from Ranging and JSpOC\\\hspace{\textwidth}
	(Predictions use only data available before epoch)}
	\label{fig:ODpredict}
\end{figure}

Figure \ref{fig:ODpredict} shows positions offsets for a ranging orbit solution produced only using
ranging data that was available before the reference epoch 
(the last ranging point used in the fit was acquired at 2015-02-24 11:45:03 UTC).  
It is compared to a high-quality post-processed GPS-derived ephemeris to show the predictive capability 
of a typical solution from the orbit determination routine.
Also shown is the position offset between the GPS-derived ephemeris and the latest JSpOC TLE 
that was available before the reference epoch, exemplifying the accuracy observed if using TLEs
for orbit prediction.

For predictions, we observe position offsets for ranging-derived and JSpOC TLEs of 0.56 and 1.82 km RMS
respectively (maximums of 1.03 and 3.84 km) when compared to the post-processed GPS-derived ephemeris.
These position offsets are quoted for the 24 hour interval following the reference epoch, the 
period of greatest interest for accurately predicting X-band passes, 
and are representative of on-orbit behaviour across the fleet. These results highlight 
the advantage of our in-house OD capability over reliance on JSpOC TLEs.

\begin{figure}[!ht]
\centering\includegraphics[width=4.5in]{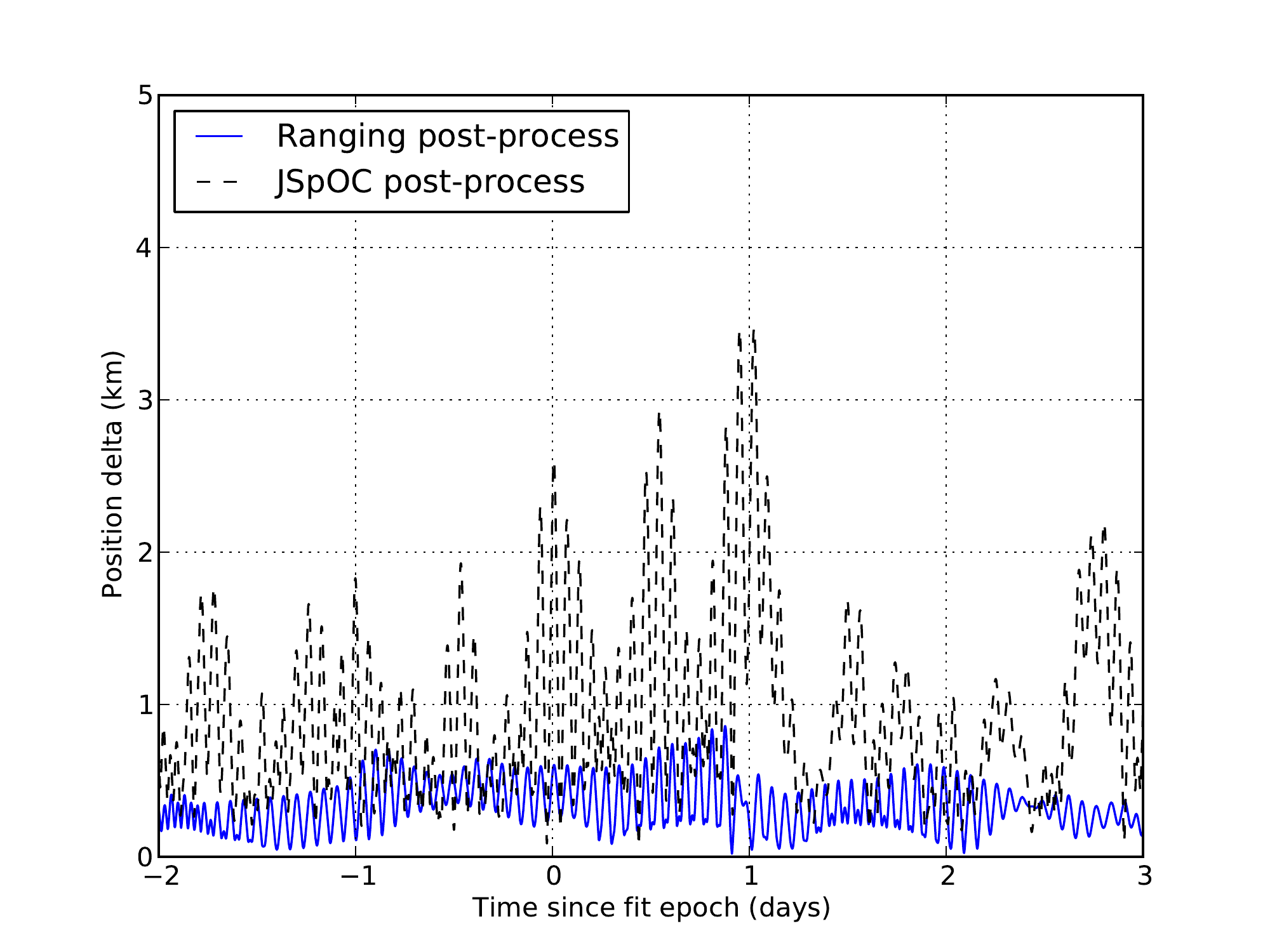}
\caption[caption]{GPS Truth vs.  Post-processed Ranging and JSpOC solutions\\\hspace{\textwidth}
(Post-processed uses all available data)}
\label{fig:ODpostprocess}
\end{figure}

\subsection{Responsible Operations}

Planet Labs' mission operators, some having previously worked in this field at NASA, 
are acutely aware of the threat that debris poses to future satellite operations.  
We care deeply about it and are determined to operate openly and responsibly in space 
to ensure sustainable and universal access to low-Earth orbit, 
which we view as a limited and shared resource.  
We have put in place several best-practice policies to limit the negative impact of our satellites.

\subsubsection{Launch Low}
Selecting orbits that have low altitudes is the single biggest factor for reducing Planet's impact on the space environment.  
Launching low reduces the time spent in orbit and ensures that we are flying below the congested 700-1000 km orbital band.  
In the unfortunate event of a collision, debris fragments will also have a short lifetime, 
further reducing the negative impact.  
Planet's model of many small, relatively cheap, and short-lived satellites 
allows us to launch into orbits with lifetimes of a few years, 
as opposed to larger satellites that prefer to maximize time in orbit.
We always ensure that our satellites are well below the UN 25 year guideline, 
and we will never launch a satellite into an orbit with a predicted lifetime greater than 25 years.

\begin{figure}[!ht]
	\centering\includegraphics[width=4.5in]{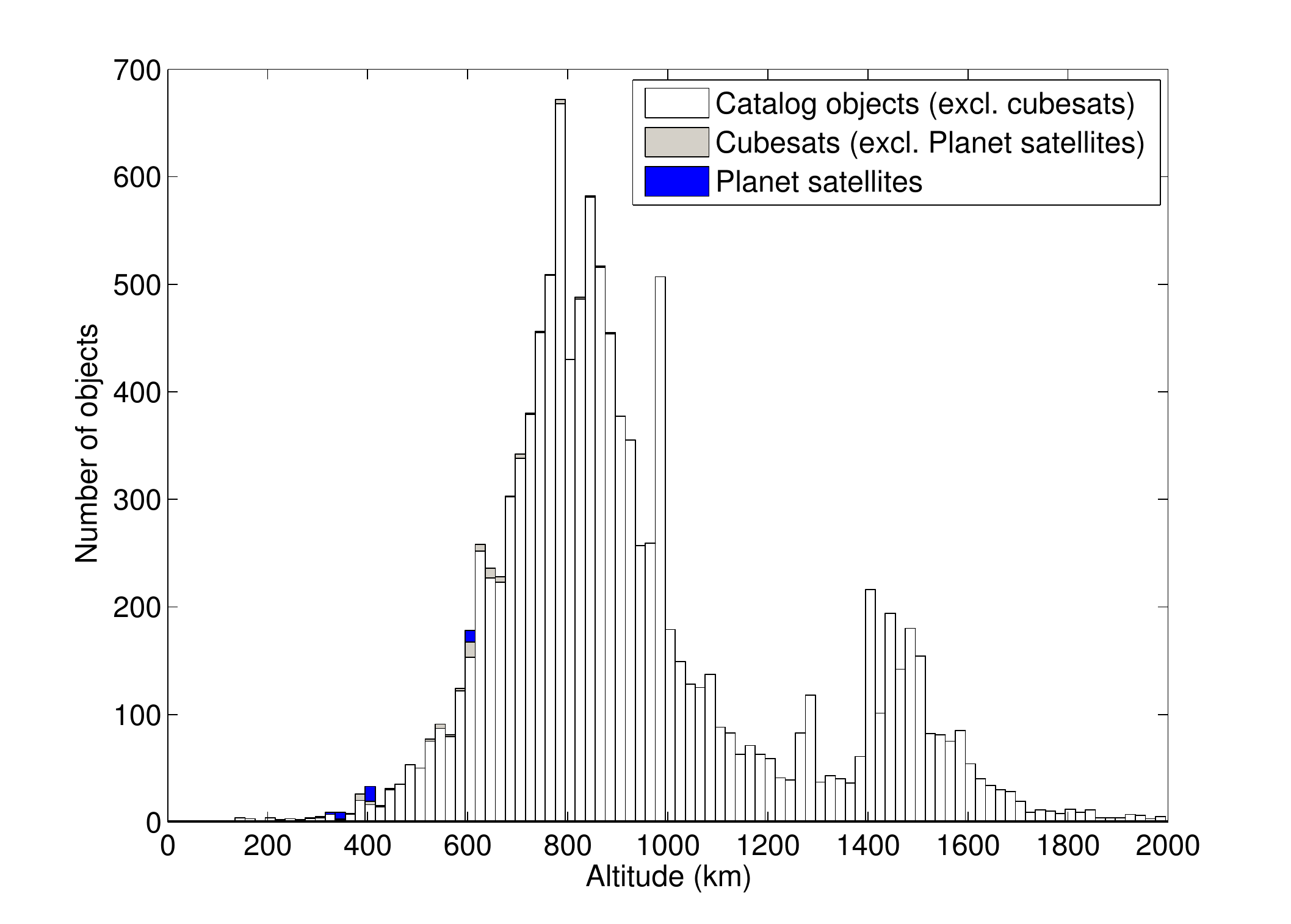}
	\caption[caption]{Semi-major axis altitude of NORAD-tracked space objects in LEO.  
	Planet Labs cubesats fly under the congested 700-1000 km region
	and exhibit lifetimes of less than 25 years.}
	\label{fig:Altitudes}
\end{figure}

\subsubsection{Open orbit data} 
The greatest limitation with current collision predictions is the poor quality of TLE-derived orbits, 
which results in many false positives.  
To help increase the accuracy of collision predictions Planet maintains an orbit catalog of its own, 
which is produced using ranging data from several ground stations 
and is publicly published hourly to \url{http://ephemerides.planet-labs.com}.  
We are a member of the Space Data Association (SDA),
which streamlines the delivery of our orbit solutions to other operators.  
In the event of a predicted close approach, 
our high-precision ephemerides are directly available to the other operators 
to help them better understand the risk and decide on a course of action.  
Planet also reports all launches and orbits to JSpOC, 
and shares tracking data to assist them in identifying our satellites upon deployment.

\subsubsection{Conjunction Response}
It is inevitable over time that high risk conjunctions will occur, 
and if the risk is deemed unacceptable we may want to take preventative actions.  
Planets Labs' satellites are not equipped with propulsion, so avoidance options are limited.  
For all conjunctions with active, maneuverable satellites, 
we will contact the operator and point them to our public-facing ephemerides server 
to ensure that they have the best-available ephemeris for our satellite.  
If the conjunction is with any other object and has a small miss distance (and similarly small uncertainties), 
then we command the Planet satellites into a minimum interaction area attitude configuration at the time of closest approach.  
Since the collision probability is a function of the combined hardbody radii of the two conjuncting objects, 
this simple attitude maneuver can result in a 30\% reduction in risk of collision with a 1 m-sized object, 
and the gains increase if the other object is smaller.  
It may also be feasible to use differential drag to avoid conjunctions, given enough lead time.  
We have developed a draft policy for differential-drag avoidance maneuvers, 
and are currently investigating the control authority and uncertainties involved before moving ahead with implementation.

\section{Differential-Drag Control}

In the second part of this paper, we present methodology and mission results from relative motion differential-drag control 
of satellites deployed in the same orbit.

\subsection{Line scanner for the Earth}

\begin{figure}[!ht]
\centering\includegraphics[width=4.0in]{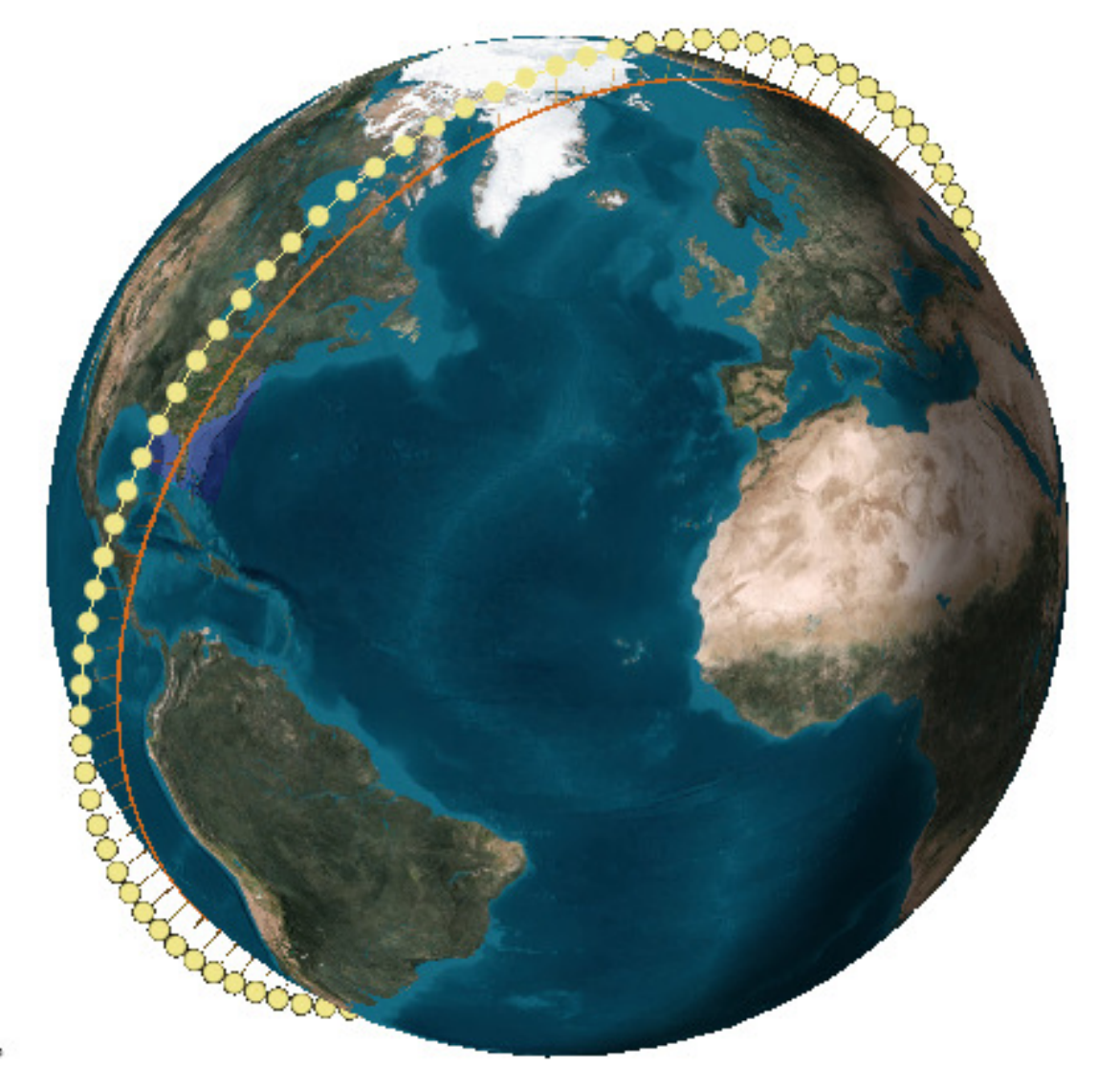}
\caption[caption]{Line scanner configuration for Planet Labs Earth imaging satellites\\\hspace{\textwidth}
Satellites are equally spaced in orbit and numbered to provide uninterrupted imaging coverage along the orbit path.
The Earth rotates beneath, yielding global coverage on a daily basis.}
\label{fig:linescannerfull}
\end{figure}

The ideal Planet Labs’ constellation is launched on a single rocket and placed into a 475 km sun synchronous orbit.  
Given perfect station-keeping, the number of satellites needed to achieve daily revisits is 105.  
If station keeping is not perfect, then there will be gaps in the coverage wherever satellites are spaced too far apart.  
This can be compensated for by additional satellites in the plane, to ensure some imaging overlap margin, 
although this needs to be accounted for in downlink budgets.  
In this way differential drag control authority can be traded directly against number of satellites (and amount of redundant data).  
To keep the cost and complexity of the fleet to a minimum it is strongly desired to minimize phasing control error.  
A constellation of evenly-spaced satellites has the additional benefit of minimizing
ground station scheduling conflicts, and therefore maximizing the amount of imagery that can be downloaded from each spacecraft.

\subsection{Maneuvering without propulsion} 

The Planet Labs satellites do not have propulsion systems, 
but since their orbits are low enough (\textless 600 km) 
we are able to use differential-drag control to achieve the desired formation.
Control is achieved by modulating the background attitude mode of the satellite 
when it is not imaging or communicating with a ground station.  
Since the satellites are highly non-spherical, different attitude modes yield different ballistic coefficients,
different rates of atmospheric decay and therefore different rates of mean motion increase.
By controlling the amount of time each satellite spends in a high-drag mode, 
one can ensure all satellites end up with the same mean motion resulting in zero relative speed.
By adjusting when these high-drag maneuvers are made, 
one can target each satellite to a desired orbit slot relative to its neighbors.
Unlike with propulsive thrusters, one is limited to controlling the along-track positioning of a satellite 
when depending on differential-drag control, since one is only effectively modulating the rate of decay due to atmospheric drag.

\begin{figure}[!ht]
\centering\includegraphics[width=3.2in]{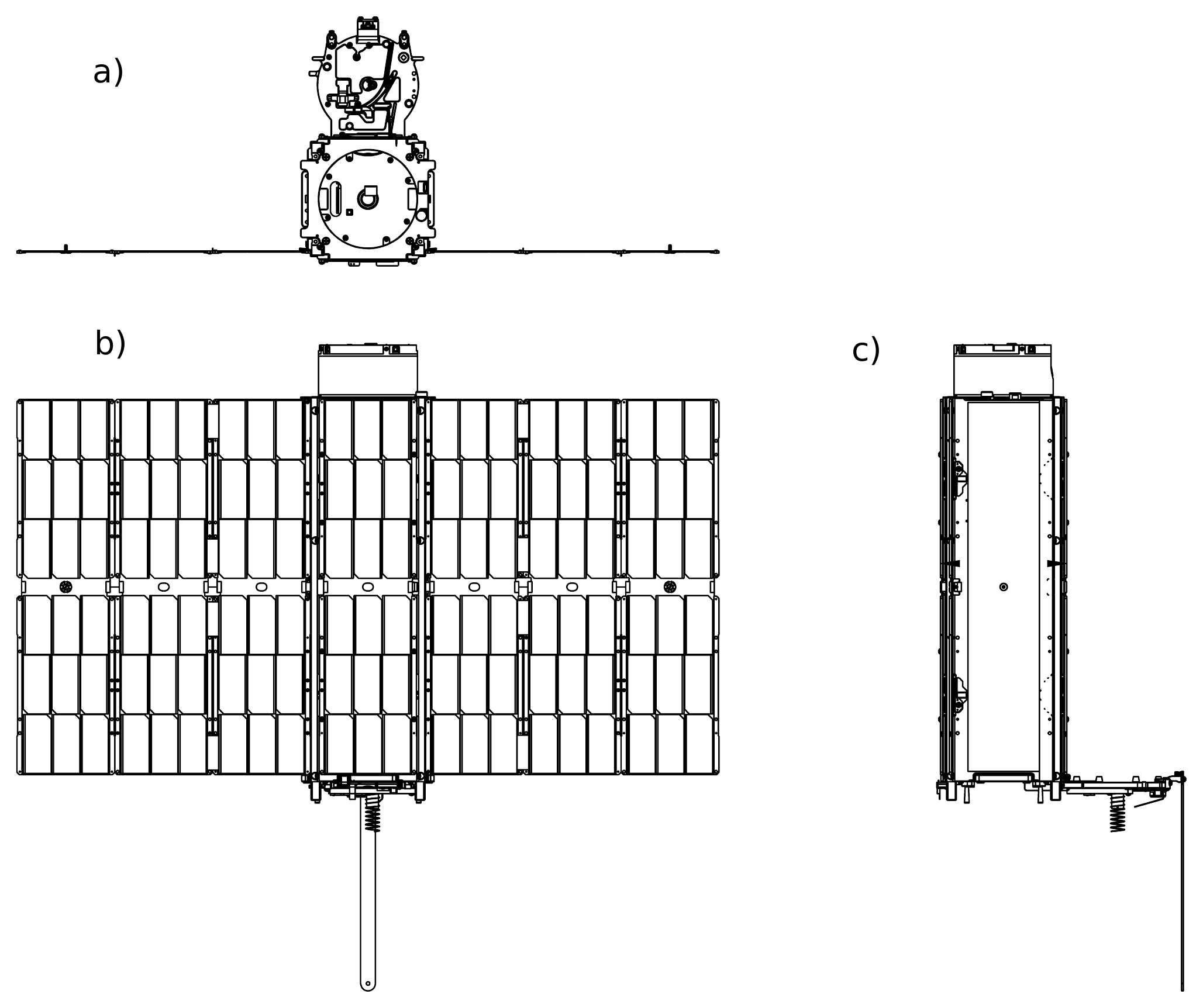}
\caption[caption]{Views of Planet Labs Dove satellite with cross-sectional areas\\\hspace{\textwidth}
a) Camera facing ($200 cm^2$)\\\hspace{\textwidth}
b) Solar-panel facing ($1950 cm^2$)\\\hspace{\textwidth}
c) Side-panel facing ($370 cm^2$)}
\label{fig:DoveDrawings}
\end{figure}

Due to operational considerations, we use two discrete attitude modes when a satellite is not occupied with imaging or downlink tasks:
a nominal low-drag mode which is preferred for power collection, and a high-drag mode afforded by the satellite's large deployed solar panels.
The nominal low-drag and high-drag modes correspond to faces b) and c) respectively of figure \ref{fig:DoveDrawings} normal to the velocity vector.
These modes result in an approximate 5:1 ratio in surface areas exposed to the "wind".
The control authority from this arrangement depends widely on altitude and atmospheric conditions, 
but ranges from $\sim1$ km/day\textsuperscript{2} in a 600 km sun-synchronous orbit 
to $\sim50$ km/day\textsuperscript{2} and greater in a sub-ISS 400 km orbit.

Attitude maneuvers are achieved using the satellite's Attitude Determination and Control System (ADCS),
consisting of a star tracker and magnetometers for estimation, and reaction wheels and magnetic torquers for control.

\subsection{Control Methodology}

We define orbit slots at equally spaced angular distances from a fixed reference satellite labelled the "leader", 
conveniently chosen to be the satellite in the lowest orbit (highest mean motion) since it is least likely to require any high-drag maneuvers.
We then create a two-component state for each satellite as its mean Earth-centered angular distance to its desired slot $\theta_i$, 
and its angular velocity relative to the leader $\dot{\theta}_i$.
The objective of the control problem is to then target a final state 
consisting of the desired orbital slots and zero relative angular velocity respectively on all satellites.  

\subsubsection{Initial dispersion}
We assume that the constellation satellites are all launched together 
and deployed with random $\Delta V$ components relative to the velocity vector.
The latter assumption ensures the initial separation of satellites in the along-track direction, 
as observed on the Planet Labs Flock 1-c launch from a Dnepr with $\pm 1$m/s along-track dispersion.
If the initial relative velocities are not sufficient to initially disperse the satellites in a timely manner,
one can command satellites to each enter high-drag mode for an appropriate amount of time\cite{alan}.

\subsubsection{Slot assignment}
The choice of which satellites to send to which orbital slots relative to the leader is made based on their initial dispersion.
We can calculate the amount of time required for each satellite to reach each slot using components of algorithm \ref{DDalgorithm}, 
then choose the vector of unique slot assignments that minimizes that total amount of time to realize the formation.
The assignment of slots needs to be performed after deployments and initial relative velocities has been observed,
and repeated if there is significant change to the availability of satellites that perform differential drag.

\subsubsection{Algorithm}
We present in algorithm \ref{DDalgorithm} a procedure to issue high-drag windows to a fleet of spacecraft
to achieve desired orbital spacing.  
We define these windows as scheduled periods of time when a spacecraft has been
commanded to maintain a high-drag attitude when not imaging the Earth or engaged in a downlink pass.
As a result of operational limitations, the control problem is non-linear since we only have two discrete attitude modes to work with, 
a nominal low-drag mode and a maneuvering high-drag mode.  
The strategy is to first calculate the required amount of high-drag time $\Delta t_{i,hd}$ for each satellite to match its mean motion
to the leader (the satellite with the lowest orbit which does not need to perform high-drag).  
One then computes the amount of time each satellite should wait in nominal low-drag mode $\Delta t_{i,wait}$ before entering this high-drag window,
such that each satellite matches the mean motion of the leader and enters its desired orbital slot simultaneously.

\begin{algorithm}
	\caption{Differential Drag Control Algorithm}
	\label{DDalgorithm}
	\begin{algorithmic}[1]
		\Procedure{Create high-drag windows}{}
		\State remove any previously issued high-drag windows
		\State compute $\theta_i$, $\dot{\theta}_i$ of each satellite $i$ relative to previous leader (mean, not osculating)
		\State compute $\ddot{\theta}$ of a satellite in high-drag vs.  low-drag mode
		\State assign leader to be satellite with greatest $\dot{\theta}$ (i.e.  lowest altitude)
		\For{each satellite $i$ that is not the leader}
		\State offset $\theta_i$, $\dot{\theta}_i$ to be relative to new leader:  
			$\theta_i=\theta_i-\theta_{leader}$, $\dot{\theta}_i=\dot{\theta}_i-\dot{\theta}_{leader}$
		\State get required high-drag duration to null relative speed with leader: 
			$\Delta t_{i,hd} = -\dfrac{\dot{\theta_i}}{\ddot{\theta}}$
		\State get angle travelled during desired high-drag window: 
			$\theta_{i,hd}=\dfrac{1}{2}\ddot{\theta}\Delta t_{i,hd}^{2}$ 
		\State get wait time to target desired slot:
			$\Delta t_{i,wait}=\dfrac{\theta_{i,hd}-\theta_{i}}{\dot{\theta_i}}$
		\State create high-drag window for satellite $i$ starting at 
			$t_{now}+\Delta t_{i,wait}$ for duration $\Delta t_{i,hd}$
		\EndFor
		\EndProcedure
	\end{algorithmic}
\end{algorithm}

The predictions for the durations $\Delta t_{i,hd}$, $\Delta t_{i,wait}$ are made with assumptions on the future
state of the atmosphere and precise attitude of the spacecraft.
The accuracy of this forecasting naturally becomes worse with time, so we rerun the algorithm on a regular cycle, 
and select an update frequency to balance between phasing accuracy, 
number of high-drag windows and availability of command upload opportunities.

\begin{figure}[!ht]
	\centering\includegraphics[width=4.5in]{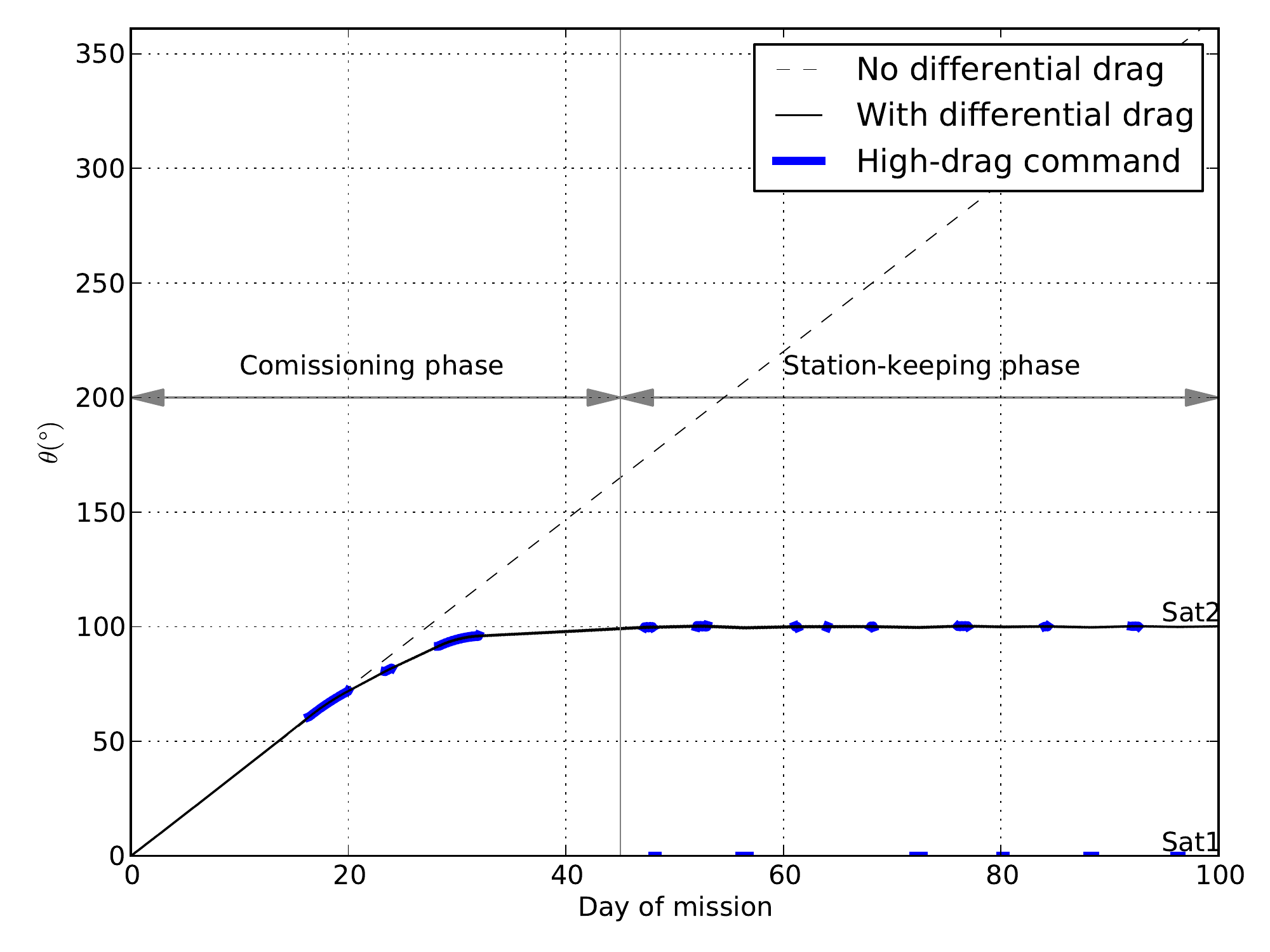}
	\caption[caption]{Simulated behaviour of two satellites performing differential drag.\\\hspace{\textwidth}
	$\theta$ is the relative Earth-centered angle between the satellites.  Thick lines represent high-drag windows}
	\label{fig:DemoDiffdrag}
\end{figure}

We present in figure \ref{fig:DemoDiffdrag} the simulated behaviour of two satellites performing differential drag
using the procedure from algorithm \ref{DDalgorithm}.
The two satellites initially remain in nominal low-drag mode due to their initial relative speed following deployment, 
then Sat2 (in a higher orbit) performs a high-drag maneuver to match mean motion with Sat1 (in the lower orbit),
and timed to do so as the former enters the desired orbit slot at 100$\degree$ from the latter.
We distinguish the commissioning phase, from deployment to first attainment of desired slot, 
from the subsequent station-keeping phase, but the same control algorithm is used throughout.

\subsection{On-orbit results}

\begin{figure}[!ht]
\centering\includegraphics[width=4.5in]{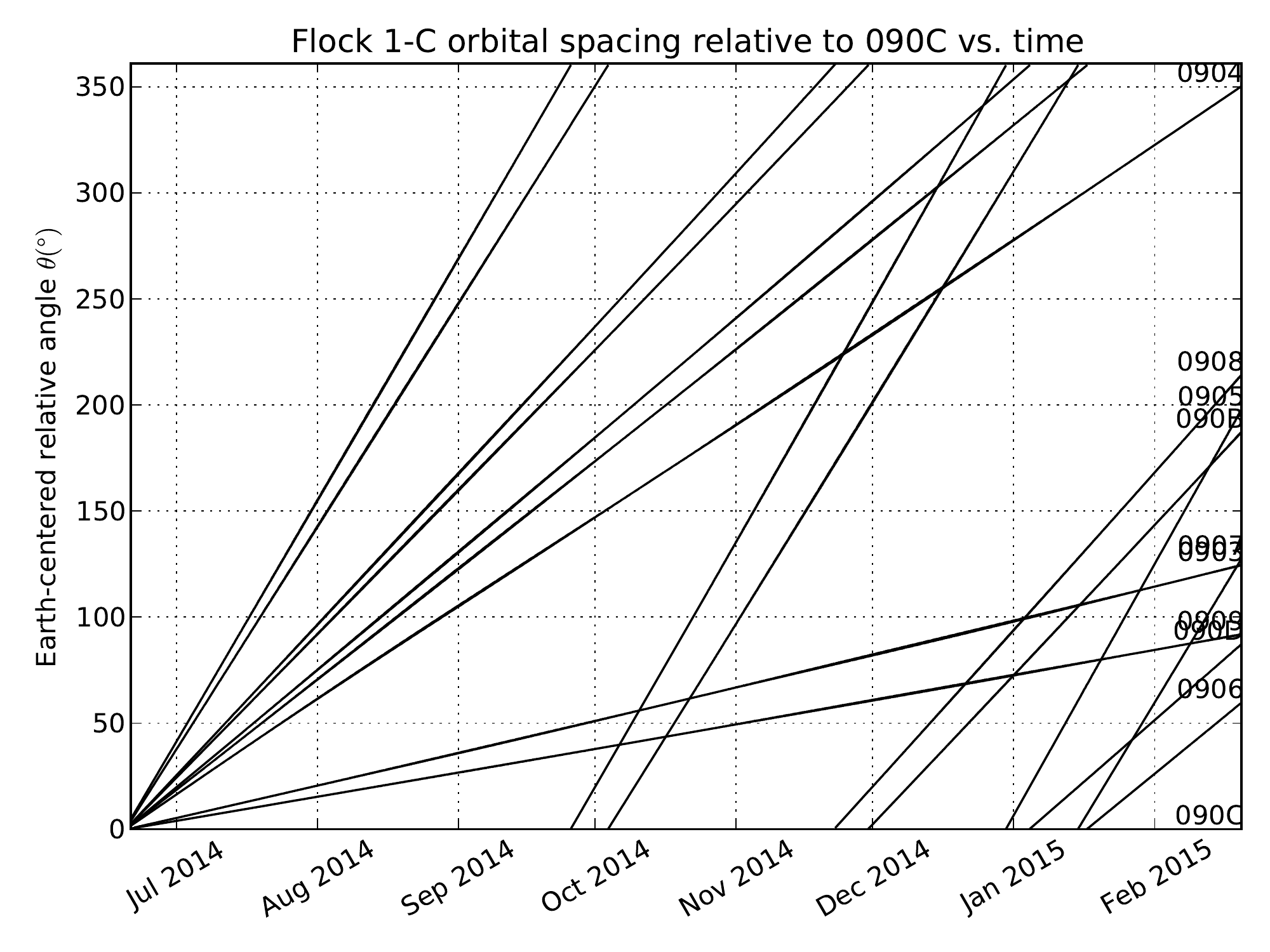}
\caption[caption]{Predicted Flock 1-C relative motion without differential drag.
Note angles wrap around at 360\degree.
Final state exhibits non-optimally placed satellites with various clumps and gaps, 
and high relative speeds.}
\label{fig:Flock1cDiffdragNoDD}
\end{figure}

\begin{figure}[!ht]
\centering\includegraphics[width=4.5in]{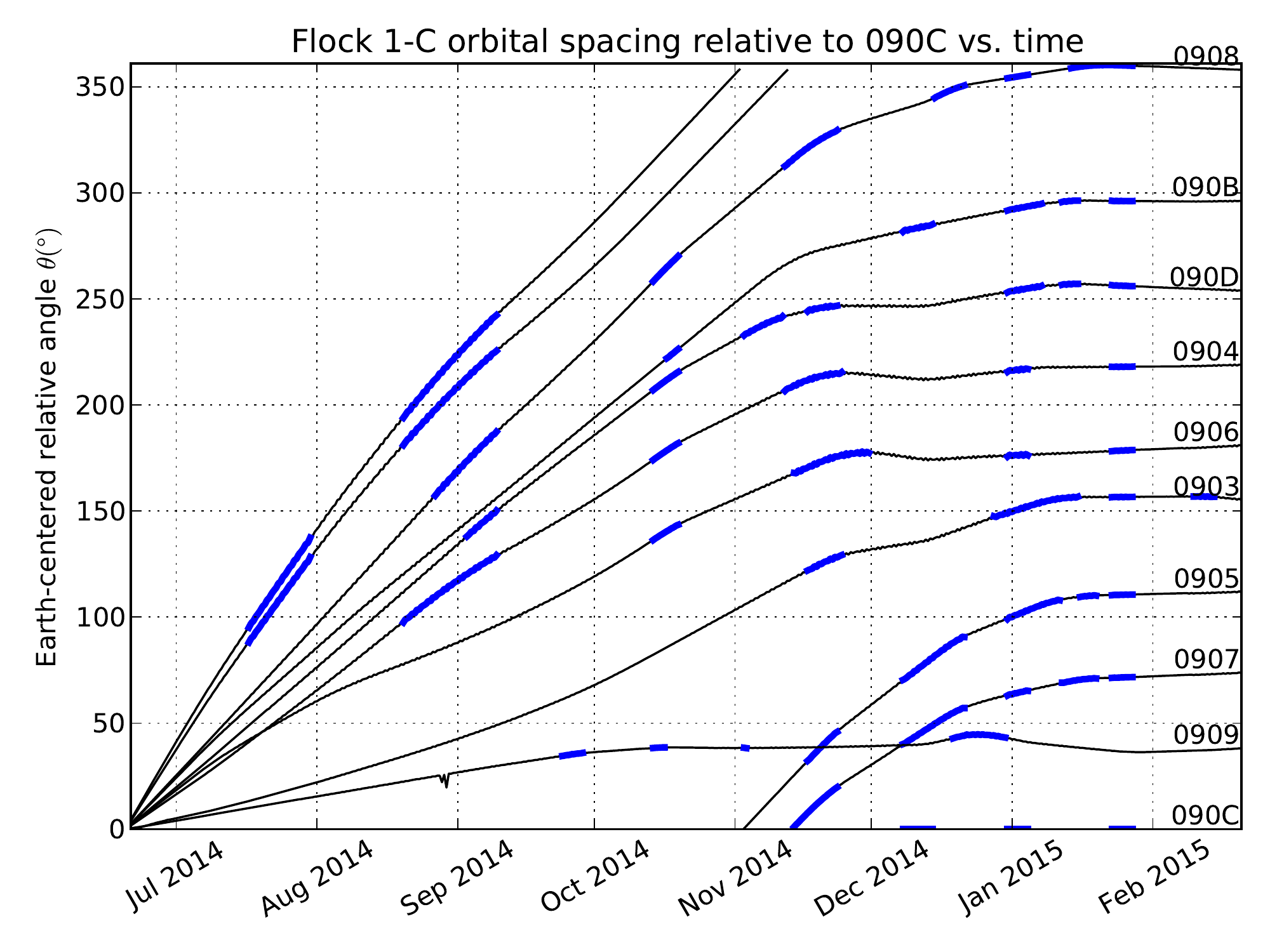}
\caption[caption]{Actual Flock 1-C relative motion achieved by using differential drag, 
plotted from acquired ranging data.
Thick lines represent executed high-drag windows. 
Final state exhibits equaly spaced satellites with very low relative speeds.}
\label{fig:Flock1cDiffdrag}
\end{figure}

We applied the differential drag control algorithm to a flock of 10 satellites in an approximately 600 km altitude sun-synchronous orbit.
Figure \ref{fig:Flock1cDiffdragNoDD} shows a prediction of relative motion of these satellites if no differential drag control were implemented,
while figure \ref{fig:Flock1cDiffdrag} shows the actual real-life behaviour of the flock from July 2014 through February 2015, 
as determined from ranging data.  
As one can see, we were successful in guiding the satellites to roughly equally-spaced slots and near-zero relative speed
by using a sequence of high-drag windows over the course of six months.
Satellites 0908 and 090C were a special case with assigned slots 2$\degree$ apart, 
to demonstrate a relatively close formation, representative of the angular distance between equally-spaced satellites on upcoming launches.

During this period, the flock was under the influence of algorithm \ref{DDalgorithm}, 
albeit with varying update frequencies due to higher priority operational constraints,
resulting in periods without any high-drag windows across the fleet.
Such delays implied that slot assignments had to be reassessed to ensure the time to realize the constellation was kept a minimum 
when differential-drag control was resumed.

\section{Conclusion}

From the very early stages of Planet Labs' conception we have identified station keeping, 
and specifically "passive" differential-drag station keeping as a desirable capability.  
In line with Planet's "Agile Aerospace", capabilities-driven engineering philosophy 
we have iteratively developed the in-house capability to perform both orbit determination 
and station keeping maneuvers.  

We have described and demonstrated successful results for orbit determination using two-way ranging data
to reliably predict orbits with sufficient quality to perform X-band passes for data downlink from Planet Labs imaging satellites.
While the current approach to orbit determination is sufficient for the high-altitude 600 km orbit, 
and most of the time in the low-altitude sub-ISS orbit, 
we recognize that further improvements can be made by incorporating short-scale attitude maneuvers
in the orbit propagator, and operationalizing the use of GPS data.

We have also detailed our approach to differential drag control and presented successful mission results 
from the phasing of 10 satellites to specific orbital slots and near-zero relative speed.
The next challenge will be applying this approach to larger fleets of 100+ satellites deployed
simultaneously in the same orbit, both for initial orbit phasing and to maintain equally spaced relative motion.

Ranging and differential drag are capabilities central to the concept of operation of a large fleet of 
nadir-imaging satellites in a single orbital plane, 
designed to apply the line scanner strategy to capture imagery of the whole world every day. 
These techniques fit naturally into Planet Labs' wider strategy for responsible and efficient operation 
of large fleets of propulsionless nanosatellites.


\bibliographystyle{AAS_publication}   
\bibliography{references}   

\end{document}